\documentstyle[12pt]{article}
\begin{document}
\newcommand{\gsim}{\mbox{\raisebox{-1.0ex}{$\stackrel{\textstyle >}
{\textstyle \sim}$ }}}
\newcommand{\lsim}{\mbox{\raisebox{-1.0ex}{$\stackrel{\textstyle <}
{\textstyle \sim}$ }}}
\newcommand{\bfx}{{\bf x}}
\newcommand{\bfy}{{\bf y}}
\newcommand{\bfr}{{\bf r}}
\newcommand{\bfk}{{\bf k}}
\newcommand{\bkp}{{\bf k'}}
\newcommand{\order}{{\cal O}}
\newcommand{\beq}{\begin{equation}}
\newcommand{\eeq}{\end{equation}}
\newcommand{\beqn}{\begin{eqnarray}}
\newcommand{\eeqn}{\end{eqnarray}}
\newcommand{\beqnn}{\begin{eqnarray*}}
\newcommand{\eeqnn}{\end{eqnarray*}}

\rightline{\large\baselineskip16pt\rm\vbox to20pt{
               \hbox{OCHA-PP-79}
               \hbox{June 1996}
\vss}}%
\vskip15mm
\begin{center}
{\large\bf Baryogenesis with vector-like quark model \\
in charge transport mechanism}
\end{center}

\begin{center}
{\large Tomoko Uesugi{\footnote{E-mail:tomoko@phys.ocha.ac.jp 
 }}, Akio Sugamoto and Azusa Yamaguchi} \\
\sl{Department of Physics, Ochanomizu University, Tokyo 112, Japan}
\end{center}
\begin{abstract} 
The electroweak baryogenesis is studied in the charge 
transport mechanism with the vector-like quark model. 
Introducing an extra vector-like up-type quark and a  
singlet Higgs scalar with the mass of the order of a few 
hundred GeV, the baryon number 
generation from the bubble wall is estimated. 
We show that this scenario is consistent with the measurement 
of the present baryon to entropy ratio of our universe, if the 
parameters are in the right region.
\end{abstract}
\vskip1cm


\baselineskip25pt
\section{Introduction}

The baryogenesis scenario in the electroweak scale was suggested in 
order to solve a number of problems in the GUT scale baryogenesis. 
Within the minimal standard model, however, this electroweak 
baryogenesis scenario faces other serious problems. 
First, CP-violation source coming only from the CKM phase is too small 
to explain the observed baryon to entropy ratio.  Second, the 
electroweak phase transition is of the weak first order one in 
the minimal standard model, although the strong first order phase 
transition is required for the baryogenesis to 
be realized.  Third, after the phase transition finishes, 
the difficult wash out problem exists.  
In order to solve these serious problems, we have to go 
beyond the standard model.

Recently, Nelson, Kaplan and Cohen proposed the electroweak 
baryogenesis scenario called the charge transport mechanism \cite{nkc}.  
If the electroweak phase transition is of the first order, bubbles 
of the broken phase appear in the symmetric phase during the 
development of the phase transition. The charge transport mechanism 
is the scenario in which quarks reflect from the expanding bubble 
wall in a manner that produces net hypercharge flux in the symmetric 
phase.  This hypercharge production is guaranteed by the CP-violating 
interactions inside the wall.  Then the produced hypercharge flux is 
converted into the baryon asymmetry through the sphaleron transition 
active in the unbroken phase outside the bubble wall.

  In this paper, we study the electroweak baryogenesis 
with the vector-like quark model.  This is a minimal 
extension of the standard model in which extra SU(2) 
singlet vector-like quarks and one extra singlet Higgs 
scalar are added to the standard model.  Historically 
this model was first introduced by Bento, Branco and Parada \cite{bbp} 
as one of the simplest spontaneous CP-violation models.
   The vector-like quark model is very attractive to the 
electroweak baryogenesis, since it can not only accommodate 
an extra CP violation source, but 
also makes the first order phase transition of the electroweak 
theory much stronger.  The baryogenesis problem with the 
vector-like quark model was studied by J.McDonald \cite{jm} 
using the spontaneous baryogenesis mechanism which is different 
from the charge transport mechanism studied in this paper.

\vspace{10mm}

\section{The vector-like quark model}
  
We consider here an extension of the standard model 
by adding extra SU(2) singlet quarks $U_L$ and $U_R$ 
as the fourth generation as well as an extra singlet 
Higgs scalar $S$. These extra singlet quarks are called 
vector-like, since they have left-right symmetric weak 
isospin and their couplings with the gauge fields become 
vector-like.

 The field content of this model is 
\beqnn
   \left( u \ d \right)^{i}_{L}, \hspace{5mm} u^{\alpha}_{R}, \hspace{5mm}
     d^{i}_{R}, \hspace{5mm} U_{L}, \hspace{5mm} \phi, \hspace{5mm} S, 
 \\    i=1,2,3, \hspace{5mm} \alpha=1,2,3,4
\eeqnn
 where $i$ and $\alpha$ are the generation indices, 
and $\phi$ denotes the standard Higgs doublet. We assume 
that the additional singlet Higgs $S$ is a complex scalar 
and all the new fields introduced, $U_L$, $U_R \equiv u^4_R$ and $S$, 
are odd under a $Z_{2}$ symmetry, whereas the ordinary 
fields are even.

Under these conditions, the $SU(2)\times U(1)\times Z_{2}$ invariant 
Yukawa couplings are given by
\beqn
    L_{Y}=-\sqrt{2}(\bar{u} \bar{d})^{i}_{L}(h_{ij}\tilde{\phi}d^{j}_{R}
           +f_{ij}\phi u^{j}_{R})-\mu\bar{U_{L}}U_{R}
           -\sqrt{2}(f_{i4}S+f^{'}_{i4}S^{*})\bar{U_{L}}u_{R}^i+h.c. \ ,
\eeqn
with additional coupling constants $\mu$, $f_{i4}$ and $f^{'}_{i4}$ 
$(i=1 \sim 3)$.

 We consider a scenario in which the extra singlet scalar, $S$,  
obtains a vacuum expectation value at the higher energy scale 
before the electroweak phase transition starts, and later the 
doublet Higgs, $\phi$, obtains the vacuum expectation value during 
the electroweak phase transition. Therefore after the electroweak 
phase transition, $S$ and $\phi$ are able to have the following 
vacuum expectation values,
\beqn
   \langle \phi \rangle =\frac{1}{\sqrt{2}} \pmatrix{ 0 \cr v \cr} \ \ 
, \ \
   \langle S \rangle = \frac{V}{\sqrt{2}} e^{i \alpha} .
\eeqn

The most general renomalizable $SU(2)\times U(1)\times Z_{2}$ 
symmetric scalar potential is as follows,
\beqn
   & & V=V_{\phi}+V_{S}+V_{\phi,S}, \nonumber \\ 
   & & V_{\phi}=\frac{\rho^{2}}{2}\phi^{\dagger}\phi+\frac{\lambda}{4}
       (\phi^{\dagger}\phi)^{2} , \nonumber \\
   & & V_{S}=S^{*}S(a_{1}+b_{1}S^{*}S)+(S^{2}+S^{*2})(a_{2}+b_{2}S^{*}S)
       +b_{3}(S^{4}+S^{*4}) , \nonumber \\
   & & V_{\phi,S}=\phi^{\dagger}\phi\left[c_{1}(S^{2}+S^{*2})+c_{2}S^{*}S
 \right], 
\eeqn  
where all the coefficients are assumed to be real. 
With the given profile of the expectation values, $v$ and $V$,  
expectation value of the phase $\alpha$ can be obtained by minimizing 
the potential with respect to $\alpha$, that is,  
\beqn
      \alpha=\frac{1}{2}\cos^{-1}\left[-\left(\frac{a_{2}+b_{2}V^{2}/2
                +c_{1}v^{2}/2}{b_{3}V^{2}}\right) \right] \label{alpha} .
\eeqn
From this eq.(\ref{alpha}) we understand that the value of $\alpha$ 
can be nonvanishing even for $v=0$ before the electroweak phase 
transition starts.  
Furthermore it can be position-dependent inside the bubble wall, 
during the electroweak phase transition is undertaking, reflecting 
the space-dependency of $v$ there. The position-dependent complex 
phase is the manifestation of the spontaneous CP 
violation in this model.  There is, however, a problem inherent 
to the spontaneous CP violation, that is,  there are 2 types among 
the 4 independent solutions satisfying eq.(\ref{alpha}), having 
different CP properties. That is, if the phase $\alpha_{+}$ is 
one of the solutions, 
the phase $\alpha_{-}$ satisfying $\sin(\alpha_{-})=-\sin(\alpha_{+})$ 
is also the solution.
Therefore, generation of the bubbles having these 
different CP phases may cancel the net hyperchage production from 
the bubble walls.  In order to avoid this difficulty, 
an explicit but tiny CP breaking may be helpful.  Following \cite{jm}, 
if we replace the factor  $(S^{2}+S^{*2})$ in the potential $V_{\phi,S}$ 
to $(e^{i\alpha_{0}}S^{2}+e^{-i\alpha_{0}}S^{*2})$, then the difference of 
the potential $\Delta V$ between two solutions of opposite signs, 
$\alpha_{+}$ and $\alpha_{-}$ is roughly 
$\Delta V \sim O(V^{4})\sin(\alpha_{0})|\sin(2\alpha_{\pm})|$.  
Let us consider ($-$) bubbles having less potential 
energy than the surrounding ($+$) bubbles by $\Delta V $.
In order for the ($-$) bubbles can grow up 
to fill the whole universe, the critical radius $r_{c}$ of the ($-$) 
bubble in the surrounding ($+$) bubbles should be less than the radius 
of the universe $r_{H}$ : 
\beqn
r_{c} \leq r_{H} \ . \label{eq:rcrh}
\eeqn 
The critical radius $r_{c}$ is the ratio of the surface tension 
$\sigma \sim O(V^{3})$ over the $\Delta V $, $r_c \sim \sigma/\Delta V 
\sim 1/(V \sin(\alpha_0)|\sin(2\alpha_{\pm})|)$. The radius of the 
universe, $r_{H}$, is related to the Hubble constant at temperature $T$ 
as follows,
\beqnn
r_{H} = H(T)^{-1} , 
\eeqnn
where
\beqnn
  H(T)\sim 20T^{2}/M_{pl} . 
\eeqnn 
Therefore eq.(\ref{eq:rcrh}) leads to  
\beqn
\sin \alpha_{0} |\sin2\alpha_{\pm}| \geq 
   \frac{20 T^{2}}{VM_{pl}} \sim 10^{-16},
\eeqn
where we have chosen  $T=100$ GeV, $V=500$ GeV and
 $M_{pl}=10^{19}$ GeV.
Therefore the tiny explicit CP breaking of $O(10^{-16})$ is enough to 
keep the ($-$) bubbles having the same CP properties, eliminating the
other ($+$) bubbles. 

The shape of the wall $v(z)$ can be obtained from the equation of 
motion for $\phi$.  Here, we assume that a wall profile can be 
determined 
dominantly from the potential $V_{\phi}$ as
\beqn
       v(z)=v_{0}\left(\frac{1+\tanh(z/\delta_{w})}{2}\right) ,
     \label{eq:wall}
\eeqn
where we consider only the dependency on the coordinate $z$, 
normal to the bubble wall, but we can get the more general 
solutions by boosting the wall 
profile (\ref{eq:wall}) in the $x$-$y$ plane.
Here $\delta_{w}$ represents the thickness of the wall.

Now, let us consider the Yukawa couplings of the up-type 
quarks.  As the weak eigen states, we take the bases in 
which mass matrix of the down-type quarks is diagonalized. 
In these bases, mass matrix of the up-type quarks can be written as
\beqn
     M= \pmatrix{m_{ij} & 0 \cr M_{4j} & \mu \cr} ,
\eeqn
where
\beqnn
     m_{ij}=f_{ij}v, \hspace{5mm} M_{4j}=(f_{4j}S+f^{'}_{4j}S^{*}) .
\eeqnn
This mass matrix can be diagonalized by the 
bi-unitary transformation, that is, by using the two 
unitary matrices $U_{L}$ and $U_{R}$ we have 
\beqn
     U^{\dagger}_{L}MU_{R}=M_{d}\equiv\pmatrix{\bar{m}_{ii} & 0 \cr 0 & \bar{M}
      \cr } ,
\eeqn
where $\bar{m}=(m_{u},m_{c},m_{t})$ and $\bar{M}$ is 
the vector-like quark mass $M_U$.
We assume that the vector-like quark and the singlet 
Higgs are heavier than the standard quarks and doublet Higgs.   

\section{Computation of the hypercharge flux} 

For simplicity, we consider here only top quark 
and the vector-like quark in the computation of the hypercharge flux. 
What we have to do is to solve the Dirac equation under the 
position-dependent wall profile and to find the reflection 
coefficients of the quarks from the wall. 
The two-generations' up-quark mass matrix is
\beqn
    M=\pmatrix{fv(z) & 0 \cr FVe^{i\alpha(z)}+F^{'}Ve^{-i\alpha(z)} & \mu
 \cr} ,
\eeqn
where $f=f_{33}$, $F=f_{43}$ and $F^{'}=f^{'}_{43}$ .
Using this $z$-dependent mass matrix, we will write the 
Dirac equation in the chiral bases.
For the stationary state ($ \psi(t,z) = \psi(z) e^{iEt} $), 
the Dirac equation reads

\beqn
   i\frac{\partial}{\partial z}\pmatrix{\psi_{1R}(z) \cr \psi_{3L}(z) \cr}
        = -\pmatrix{E & M^{\dagger} \cr M & -E \cr }\pmatrix{\psi_{1R}(z) 
                                 \cr \psi_{3L}(z) \cr} \nonumber  \\
   i\frac{\partial}{\partial z}\pmatrix{\psi_{4L}(z) \cr \psi_{2R}(z) \cr}
        = -\pmatrix{E & M \cr M^{\dagger} & -E \cr }\pmatrix{\psi_{4L}(z)
                                 \cr \psi_{2R}(z) \cr} 
\eeqn   
where $\psi_{L(R)}(z)$ has two components, corresponding to the top and
the vector-like quarks.
 
We have solved this Dirac equation numerically as a system 
of ordinary differential equations with respect to $z$, without 
any approximations.  
The reflection coefficient from the right-handed $j$-quark 
to the left-handed $i$-quark, $R^{ij}(E)$, and from the 
left-handed $j$-quark to the right-handed $i$-quark, $\bar{R}^{ij}(E)$, 
are given by
\beqn
  R^{ij}(E)&=&\frac{\psi^{i}_{3L}(z_0)}{\psi^{j}_{1R}(z_0)} \\
  \bar{R}^{ij}(E)&=&\frac{\psi^{i}_{2R}(z_0)}{\psi^{j}_{4L}(z_0)} 
\eeqn
where $i$ and $j$ represent the generations, $E$ is the 
incident energy and $z_0$ denotes the position of the boundary 
between the wall and the symmetric phase.
Using these reflection coefficients, the hypercharge flux can be written 
as
\beqn
  f_{Y}&=&\int^{\infty}_{0}dk_{L} \int^{\infty}_{0} k_{T}dk_{T} 
          \left(\frac{1}{4\pi^{2} \gamma }\right) \nonumber \\
    & &  \sum_{i,j} \left[(\Delta Y)_{j\to i}|R_{ij}(k_{L})|^{2}+
              (\Delta Y)_{j\to i}|\bar{R}_{ij}(k_{L})|^{2} \right] 
         \nonumber \\
    & & \times \left(f^{s}(E^{s})_j-f^{b}(E^{b})_i \right) ,
\eeqn
with
\beqn 
    f^{s}(E)_i&=&\frac{k_{L}}{\sqrt{k_{L}^{2}+k_{T}^{2}+M^{s2}_i}
                       \left(1+\exp(E^{s}/T)\right)} ,\nonumber \\
    f^{b}(E)_i&=&\frac{k_{L}}{\sqrt{k_{L}^{2}+k_{T}^{2}+M^{b2}_i}
                       \left(1+\exp(E^{b}/T)\right)} , \nonumber \\
    & & \ \ E^{s}=\gamma \left(\sqrt{k_{L}^{2}+k_{T}^{2}+M^{s2}_i}
                            -v_{w}k_{L}\right) ,  \nonumber \\
    & & \ \ E^{b}=\gamma \left(\sqrt{k_{L}^{2}+k_{T}^{2}+M^{b2}_i}
                            +v_{w}k_{L}\right) ,
\eeqn
where $(\Delta Y)_{j\to i}$  represents the change of the 
hypercharge from $j$-th quark to the $i$-th quark, $\gamma$ is 
the $\gamma$-factor $1/\sqrt{1-v_w^2}$ corresponding to the 
wall velocity $v_w$, $k_T$ and $k_L$ are transverse and 
longitudinal components of the momentum in the wall rest 
frame, and $E^s$ ($E^b$) and $M^s$ ($M^b$) are, respectively, 
the energy and the diagonalized mass of quark in the thermal 
frame of the symmetric (broken) phase. 
The integral over $k_{T}$ can be done analytically.  Then we have
\beqn
   f_{Y}&=&\int^{\infty}_{0}dk_{L}\left(\frac{k_{L}}{4\pi^{2} \gamma^{2}}
            \right)
            \sum_{i,j} \Biggl[(\Delta Y)_{j\to i}|R_{ij}(k_{L})|^{2}
         + (\Delta Y)_{j\to i}|\bar{R}_{ij}(k_{L})|^{2} \Biggr] 
          \nonumber \\
        & &  \times \Biggl[\log\left(1+\exp\left[-\gamma(k_{L}-v_{w}
            \sqrt{k_{L}^{2}-M^{s2}_j})/T \right] \right) 
          \nonumber \\ 
        & &  \hspace{10mm} -\log\left(1+\exp\left[-\gamma(k_{L}+v_{w}
            \sqrt{k_{L}^{2}-M^{b2}_i})/T \right] \right)\Biggr] .
\eeqn
We have to give the phase transition temperature and  
the masses of the quarks in the broken phase for the calculation of 
the hypercharge flux.
We set the phase transition temperature $T=100$ GeV, and the mass 
of the top quark $m_t =174$ GeV.
The wall width is considered roughly the order of the phase transition 
temperature, $\delta_{w}^{-1} \sim T \sim 100$ GeV, 
but we have estimated here the hypercharge flux for the various 
choices of the wall width.
The experimental constraints for the mass of the vector-like quark is 
not so strict \cite{bbr}.      
We have tried the cases of $M_U=500$ GeV and $300$ GeV.
Most of the parameters in the scalar potential $V$ in eq.(3) are 
outside of the experimental verification.  Therefore we can choose 
freely $a$ and $b$ with a restriction of $a+b=1$ and determine the 
position-dependent $\alpha$ as $\cos2\alpha = a + b(v(z)/v_{0})^2 $ .  
The restriction above means the disappearance of  the spontaneous CP 
violation phase after the electroweak phase transition ends.  
We also introduce $m_{com}$ and $h$ defined by $m_{com} =  (F+F^{'})V$ 
and $h =  F^{'}/F$ .  Then, in the case of $M_{U}=500$ GeV,  
$m_{com}=300$ GeV, and $v_{w}=0.5$, we have $f_{Y}= O(10^{-6 \sim -7})$ 
for the various choices of 
$(b, h) = (1.0 \sim 2.0, 0.01 \sim 0.5)$.  
The value $b = 2.0$ corresponds to the case that the complex phase moves 
from zero to $\pi/2$ inside the wall. If the value $h$ becomes smaller, 
the imaginary part becomes larger than the real part in the off-diagonal 
term. The value $h=1.0$ corresponds to the real off-diagonal term.
    
If we change the off-diagonal value, $m_{com}$, of the 
mass matrix, we have obtained the following results for 
$M_{U} = 300$ GeV, $v_{w}=0.1$, $h=0.01$, and $b=2.0$ : 
$f_{Y} = O(10^{-8 \sim -6})$ corresponding to the values of 
$m_{com}= 0.0 \sim 1.2 $. (see Fig.1) 

 Fig.2 shows the wall width dependence of the hypercharge flux in the case 
of $M_U = 300 GeV$ and $M_U = 500 GeV$. We can see that the hypercharge flux 
takes the larger value for the thinner wall.

\section{Generation of the baryon asymmetry from the hypercharge flux}

 Following the method of Nelson, Kaplan and Cohen \cite{nkc}, 
we compute the partial derivative of the free energy with respect 
to the baryon number.
 If the time scale of the baryon number violating sphaleron 
transition is much longer than that of the weak interactions 
between reflected quarks, $10^{-26} \ [s]$, we will be able to think 
the process of the sphaleron transition is not in the thermal 
equilibrium within the time scale of the weak interactions.  
Then, the baryon number and the lepton number will conserve 
during the time scale. 
 In that situation we can introduce the chemical potentials 
corresponding to hypercharge $Y/2$, charge $Q$, baryon number 
$B$, $B-L$ ($L$ : lepton number), and 
$B^{'}=(B_4+B_3)/2-(B_2+B_1)/2$, where $B_i$ is the baryon number 
of the i-th generation.  We have neglected the mixing between the 
lighter (1st and 2nd) generations and the heavier (3rd and 4th) 
generations, but this is a quite reasonable assumption.  
 Using these chemical potentials, we can obtain the relation 
between the chemical potential of the baryon number $\mu_{B}$ 
and the hypercharge density $\rho_{Y/2}$~:
\beqn
 \rho_{Y/2} &=& \frac{\mu_{B}}{24\beta+1}
                     \left[-2(24\beta+\frac{8}{3})(8\beta+1)
                     +(24\beta+1)(16\beta+1)\right] \frac{T^2}{3} 
                    \nonumber \\
           &\sim& - \frac{13}{9}\mu_B T^2 \ ,
\eeqn
where
\beqn
      \beta=\frac{1}{4 \pi^2}\int^{\infty}_0 dx
            \frac{x^2}{1+\cosh\sqrt{x^2+M_{U}^2}} \ .
\eeqn
One can easily see that the $\beta$ is small enough to 
be neglected as long as we set the mass of the vector-like quark 
$M_U$ at a few hundred GeV .
Then the net baryon number production is
\beqn
  \rho_B &=& \frac{\Gamma_{sph}}{T}\int dt \frac{\partial F}{\partial B}
            \nonumber \\
         &=& \frac{9 \Gamma_{sph}}{13 T^3} \int^{z/v_w}_{-\infty} dt
             \rho_{Y/2}(z-v_w t) \ , 
\eeqn
where $\rho_{Y/2}(x)$ denotes the hypercharge density at 
the position which is located at the distance $x$ from the bubble wall.
$\Gamma_{sph} = k(\alpha_W T)^4$ is the sphaleron transition rate 
of the symmetric phase,  where $\alpha_W=g^2/4 \pi$, and $k$ is a 
numerical factor which has been estimated by J.Ambj\o rn et al. 
about $0.1 \sim 1.0$ \cite{amb}.
The integral over $t$ can be estimated approximately by the 
transport time of the reflected quarks $\tau_T$ as follows:
\beqn
  \int^{z/v_w}_{-\infty} dt \rho_{Y/2}(z-v_w t)&=&
             \frac{1}{v_w}\int^{\infty}_0 dx \rho_{Y/2}(x) 
             \nonumber \\
             &\sim& \frac{f_Y \tau_T}{v_w} \ ,
\eeqn
and the baryon to entropy ratio becomes
\beqn
  \frac{\rho_B}{s}&=& \frac{9}{13} \frac{k\alpha_W^4 T}{s} 
                    \frac{f_Y \tau_T}{v_w} \nonumber \\
                  &\sim& 10^{-8} k \frac{(\tau_T T)}{v_w}(f_Y/T^3) .
\eeqn
 
 Then to explain the present baryon to entropy ratio, we need 
$f_Y \sim 10^{-5 \sim -6} T^3$, since the typical transport time 
of the weak scale is roughly $O(10^{1 \sim 3}T^{-1})$. The solutions
of $f_Y \sim 10^{-5 \sim -6} T^3$ have been found in the charge 
transport scenario with the vector-like quark model in the last
section, so that the scenario can be a good candidate of the model 
of the baryogenesis.

\section{Conclusion}
In this paper, the charge transport scenario of the baryogenesis
has been studied using the vector-like quark model.  We have
adopted the model in which extra left-right symmetric, $SU(2)$
singlet up-type quarks $U_L$ and $U_R$, and a $SU(2)$ singlet
Higgs scalar $S$  are added to the standard model. The possible
Higgs potential for the ordinary doublet Higgs $\phi$ and the
singlet $S$ admits the spontaneous CP violation in which the
phase $\alpha$ of $S$ takes the non-vanishing vacuum expectation
value. During the first order phase transition bubbles of the
broken phase are generated, and the phase $\alpha$ becomes position
dependent inside the bubble walls. A tiny but explicit CP violation
should be, however, incorporated in order to avoid the difficulty
of having two types bubbles. In the presence of this explicit CP
breaking, one type of the bubbles creating the baryon number
through the charge transport mechanism, remains, whereas the other
type of the bubbles decreasing it, disappears.  

By solving the Dirac equation numerically without any approximations, 
we have estimated the rate of the hyperchage production from the 
bubble walls, in which the presence of the position-dependent CP 
phase in the mass mixing 
between the top quark and the extra up-type one is important. 
The evaluated hypercharge flux from the wall $f_Y$ depends 
on the various unknown
parameters.  However, with the top quark mass of $174$ GeV, the
vector-like quark mass of $500$ GeV or $300$ GeV, and the 
phase transition temperature of $100$ GeV, we have solutions giving 
$f_Y \sim 10^{-6}$ for the value of the wall width $\delta_W =
(100 GeV)^{-1}$.
It is the reasonable value for reproducing the baryon to entropy
ratio of $10^{-10 \sim -11}$.  For the transition of the hypercharge
to the baryon number, the usual discussion has been extended so as to
incorporate the extra vector-like up-type quark and the singlet Higgs,
in which the thermal equilibrium by the weak interactions is violated
by the sphaleron transition.

The phase transition dynamics when singlet Higgs is introduced has not 
been given here.  Some of the results can be found
by using the 1-loop approximation and the high temperature 
expansion \cite{ah} on how the first order phase transition becomes
stronger in our case \cite{azusa}, but the more detailed study 
should be required.
In the real development of the first order phase transition, fusion
effect of the bubbles including the temporal change of the expanding
velocity of the bubbles, should be taken into account \cite{as}. 
This kind of analysis is difficult, but very important.  Also the
detailed analysis should be given on the disappearance of the unwanted
bubbles having the different CP properties.  In the analysis of
estimating the final value for the baryon to entropy ratio,  we
need to know the exact value of the sphaleron transition rate, or
the coefficient $k$ in eq.(21), as well as the transport time
$\tau_T$ for the various quarks. Improvement on the estimation of
these parameters are also necessary.

\vskip1cm

\centerline{\bf Acknowledgment}
We would like to thank J. Arafune, K. Yamamoto and K. Funakubo
for their advice. We are also grateful to I. Watanabe and 
H. Aoki for their useful discussions. 

This work is supported in part by the Grant-in-Aid for Scientific Research 
from the Ministry of Education, Science and Culture (No.08640357). 


\newpage
{\Large \bf Figure Captions}

\begin{enumerate}
\item{Fig.1:} The wall width dependence of the hypercharge flux in 
the case of $M_U=500$ GeV and $m_{com}=300$ GeV for $b=2.0$, $h=0.01$, 
$v_w=0.1$ and $\delta_w=T^{-1}$ (solid line). The case of $M_U=300$ GeV 
and $m_{com}=100$ GeV is also shown (dashed line).

\item{Fig.2:} The hypercharge flux as a function of 
$m_{com}=0.0 \sim 1.2$ GeV for $M_U = 300$ GeV. Other parameters are 
the same as in Fig.1.

\end{enumerate}
\end{document}